\documentclass[a4paper]{article}

\usepackage[left = 3.3 cm, right = 3.3 cm]{geometry}
\usepackage[T1]{fontenc} 
\usepackage[utf8]{inputenc}
\usepackage{lmodern}
\usepackage{bm} 
\usepackage{amsmath, amsfonts, amssymb}
\usepackage{graphicx}
\usepackage{hyperref}
\usepackage{upgreek}
\usepackage[super, sort&compress]{natbib}
\usepackage{authblk}
\usepackage{todonotes, soul, xcolor}
\usepackage{lineno}
\usepackage{setspace}

\author[1, $\dagger$]{Andr\'{a}s P\'{a}link\'{a}s}
\author[1, $\dagger$]{Kriszti\'{a}n M\'{a}rity}
\author[1, 3]{Konr\'{a}d Kandrai}
\author[1, 2]{Zolt\'{a}n Tajkov}
\author[2, 4]{Martin Gmitra}
\author[1]{P\'{e}ter Vancs\'{o}}
\author[1]{Levente Tapaszt\'{o}}
\author[1, *]{P\'{e}ter Nemes-Incze}

\affil[1]{Hungarian Research Network, Centre for Energy Research, Institute of Technical Physics and Materials Science, 1121 Budapest, Hungary}
\affil[2]{Centre of Low Temperature Physics, Institute of Experimental Physics, Slovak Academy of Sciences, Košice SK-04001, Slovakia}
\affil[3]{Department of Physics, Institute of Physics, Budapest University of Technology and Economics, M\H{u}egyetem rkp. 3., H-1111 Budapest, Hungary}
\affil[4]{Institute of Physics, Pavol Jozef Šafárik University in Košice, SK-04001 Košice, Slovakia}
\affil[$\dagger$]{\small \emph{These authors contributed equally}}
\affil[*]{\small \emph{corresponding author, email: nemes.incze.peter@ek.hun-ren.hu}}

\date{}

\title{Identification of graphite with perfect rhombohedral stacking by electronic Raman scattering}

\begin{document}

\maketitle

\doublespacing


\textbf{
	Rhombohedral graphite (RG) shows strong correlations in its topological flat band and is pivotal for exploring emergent, correlated electronic phenomena.
	One key advantage is the enhancement of electronic interactions with the increase in the number of rhombohedrally stacked graphene layers.
	Increasing thickness also leads to an exponential increase in the number of stacking faults, necessitating a precise method to identify flawless rhombohedral stacking.
	Overcoming this challenge is difficult because the established technique for stacking sequence identification, based on the Raman 2D peak, fails in thick RG samples.
	We demonstrate that the strong layer dependence of the band structure can be harnessed to identify RG without stacking faults, or alternatively, to detect their presence.
	For thicknesses ranging from 3 to 12 layers, we show that each perfect RG structure presents distinctive peak positions in electronic Raman scattering (ERS).
	This measurement can be carried out using a conventional confocal Raman spectrometer at room temperature, using visible excitation wavelengths.
	Consequently, this overcomes the identification challenge by providing a simple and fast optical measurement technique, thereby helping to establish RG as a platform for studying strong correlations in one of the simplest crystals possible.
}

\section*{Introduction}

Graphene-based electron systems are recognized for their simplicity and versatility in probing emergent, strongly-correlated electronic phenomena~\cite{Cao2018-hm,Cao2018-qw,Andrei2020-hh}.
Rhombohedral graphite (RG) has recently gained prominence~\cite{Myhro2018-me,Zhou2021-wf,Zhou2021-of,Lee2022-cu,Shi2020-bv,Hagymasi2022-hg,Han2024-eb,Liu2024-gh,Lu2024-gi,Han2023-ht,Zhou2024-kr} as a platform for such studies, surpassing even the "magic angle" twisted bilayers~\cite{Andrei2020-hh} in simplicity due to its inherent lack of twist angle disorder~\cite{Uri2020-az}.
The simplest forms of rhombohedrally stacked graphene layers: Bernal bilayer and "ABC" trilayer have been studied almost since the discovery of graphene~\cite{Novoselov2006-pk}.
Recently, unconventional superconductivity~\cite{Zhou2021-of} and Stoner type spin and valley magnetism~\cite{Zhou2021-wf} were discovered in trilayer.
The growing interest in thicker samples~\cite{Liu2024-gh,Lu2024-gi,Han2023-ht,Han2024-eb} is hardly surprising, as the strength of interactions increases with the addition of more rhombohedrally stacked graphene layers~\cite{Xu2012-bw,Pamuk2017-mj}.
In pentalayers, for instance, fractionalization of the Hall resistance has been reported even without an external magnetic field~\cite{Lu2024-gi}, while in four, five and seven layer samples a correlated insulator state at charge neutrality was demonstrated~\cite{Liu2024-gh,Han2024-eb,Zhou2024-kr}.
Applying a perpendicular displacement field in these systems yields transport measurements that indicate the presence of Chern insulators~\cite{Han2024-eb}.
Thicker RG crystals exhibit signs of competing correlated ground states, further emphasizing the material's electronic complexity.
Signatures of such states, characterized through transport measurements~\cite{Shi2020-bv} and scanning tunnelling microscopy (STM)~\cite{Hagymasi2022-hg}, highlight the emergent electronic properties inherent in this simple crystal.

To explore the rich, emergent electronic structure of the RG flat band, accurate identification of rhombohedrally stacked graphene layers, without stacking faults is of paramount importance.
Typically, RG identification hinges on confocal Raman spectroscopic mapping, where the 2D peak's shape is a key differentiator from the hexagonal phase~\cite{Torche2017-sj,Yang2019-yx}.
The peak width or the integrated intensity ratio of the two halves of the peak (below and above the midpoint)~\cite{Yang2019-yx} are used to indicate partial or full rhombohedral stacking.
For 3 and 4-layer structures, both the M peak~\cite{Cong2011-lx,Nguyen2014-nu} and the 2D peak shape~\cite{Nguyen2014-nu,Wirth2022-yl} serve as reliable indicators of the exact stacking configuration.
However, as the number of layers ($N$) increases, the distinctiveness of both peaks' shapes diminishes across different stacking configurations~\cite{Nguyen2014-nu} (see  section S1 and S2 of the Supplementary Information).
As an example, in Fig.~\ref{fig:problem}a we present measurements of the 2D peak from a graphite flake with rhombohedral stacking, in regions with a layer thickness of 7, 8 and 10.
At first glance, the peaks are indistinguishable, with the only variation being attributable to measurement noise.
A more quantitative analysis by the integrated intensity ratio of the lower and upper sides of the 2D peak~\cite{Yang2019-yx} (within the regions marked by dashed lines on Fig.~\ref{fig:problem}a) fails to adequately distinguish between the 7, 8 and 10 layer thicknesses (see Fig.~\ref{fig:problem}).
Another issue when dealing with RG is that in thicker flakes stacking faults can not be identified relying solely on the 2D peak shape, that could lead to severe inconsistencies in later measurements.
While scattering infrared near-field microscopy is also employed to identify RG domains~\cite{Han2024-eb,Liu2024-gh,Wirth2022-yl,Gao2020-bj,Beitner2023-wo}, a definitive method to distinguish thick, flawless crystals from those with stacking faults is yet to be established.
The lack of a straightforward and rapid technique to verify defect-free rhombohedral stacking in flakes with $N \geq 5$ significantly hinders work with thicker RG samples.

\begin{figure}[!ht]
	\begin{center}
	\includegraphics[width = 0.5 \textwidth]{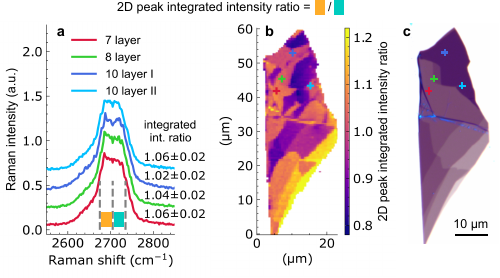}
	\caption{\textbf{Similarity of 2D peak shapes for thick RG.}
		\textbf{(a)} Example Raman spectra of the 2D peak, from areas of a graphite flake with predominant rhombohedral stacking.
		The spectra are selected from areas with 7, 8 and 10 graphene layers.
		Dashed lines show the lower and upper Raman shift values, used for calculating the integrated intensity ratio map shown in (b).
		Numbers to the right of the spectra are the integrated intensity ratio of the measurements.
		Errors stem from the local variability within the map in (b).
		\textbf{(b)} Map of the integrated Raman intensity in the range 2675 to 2705 cm$^{-1}$ divided by the integrated intensity in the range 2705 to 2735 cm$^{-1}$, as introduced by Yang et al~\cite{Yang2019-yx}.
		Larger values correspond to more prominent rhombohedral stacking.
		\textbf{(c)} Optical microscopy image of the flake.
		Positions of selected spectra shown in (a), marked by "+" signs of corresponding colour.
		Raman spectra are measured, using 532 nm excitation.
	}
	\label{fig:problem}
	\end{center}
\end{figure}

Here, we demonstrate that crystals of rhombohedral graphite (RG) devoid of stacking faults can be identified through electronic Raman scattering (ERS) measurements.
This identification relies upon independently establishing the number of layers ($N$) by optical contrast or atomic force microscopy (AFM).
Additionally, the presence of stacking faults can be easily detected through a mismatch between the layer count and the ERS peak positions.
In ERS, the inelastic scattering of a photon leaves behind an electron-hole excitation in the crystal, instead of a phonon.
The energy of the electron-hole excitation measured by the ERS signal reflects the specific DOS of the electrons in RG~\cite{Kashuba2009-oa,Garcia-Ruiz2019-dp,McEllistrim2023-op}.
This raises the possibility to accurately distinguish few layer thick RG samples, owing to their unique set of band edges which are specific to the layer number and stacking sequence.
To achieve this, two crucial elements are required: an accurate method for measuring flake thickness and a measurement of the ERS peak positions~\cite{Garcia-Ruiz2019-dp, McEllistrim2023-op}.
Atomic Force Microscopy (AFM) can provide an exact determination of thickness~\cite{Nemes-Incze2008-la}, while the precision of the ERS measurement stems from the strong dependence of the band structure on the number of layers~\cite{Xu2012-bw, Garcia-Ruiz2019-dp, McEllistrim2023-op}.
We reveal that the positions of the ERS peaks can effectively distinguish RG domains, with defect free stacking, applicable to structures with up to 12 layers and beyond.
Furthermore, in contrast to the 2D peak shape, the ERS signal exhibits no dispersion with the chosen excitation wavelength (see section S3 of the Supplementary Information).
This lack of dispersion simplifies the use of the ERS signal and facilitates comparison between different measurements.
A comprehensive, step-by-step description of the identification processes is provided in the Supplementary Information (S4) of this paper.

\section*{Results}

The 2D peak shape is the most widely used feature to distinguish hexagonal and rhombohedral stackings of graphite.
Its sensitivity to stacking configurations originates from variations in the band structure near the K points, which is sampled by the double resonant Raman process at the energies determined by the excitation laser~\cite{Torche2017-sj}.
However, the bands at these energies, quickly converge to the bulk values with increasing layer number.
This leads to the observation that, when relying solely on phonon modes, Raman spectroscopy struggles to accurately distinguish between RG of various layer thicknesses (see Supplementary Information S1) and as we show later, can not unambiguously identify the presence of stacking faults for $N > 5$.
The insufficient distinctiveness of the 2D peak is shown in Fig.~\ref{fig:problem}a.
A much stronger dependence on the layer number is displayed by a set of band edges in few layer RG~\cite{Henni2016-yc,Garcia-Ruiz2019-dp,McEllistrim2023-op}.
These band edges show up as peaks in the density of states (DOS)~\cite{McEllistrim2023-op} and their energy separation is a unique fingerprint of the RG thickness (see Fig.~\ref{fig:ers_extracting}a).
However, directly probing the DOS, for example in STM, is very time consuming, hence it is not suitable for the much needed quick characterization of RG.
By contrast, electronic Raman scattering offers a fast and versatile method that probes the excitation spectra of the electrons~\cite{Kashuba2009-oa,Riccardi2016-ir,Henni2016-yc}, thus it can provide a direct fingerprint of the layer number, by measuring the spectral features of electron-hole excitations in the vicinity of the band edges (DOS peaks). 
The ERS peaks of various stacking configurations were recently calculated by Garc\'{i}a-Ruiz and McEllistrim et al~\cite{Garcia-Ruiz2019-dp,McEllistrim2023-op}, but experimental investigation has been limited to magneto-Raman measurements~\cite{Henni2016-yc}.

\begin{figure}[!ht]
	\begin{center}
	\includegraphics[width = 1 \textwidth]{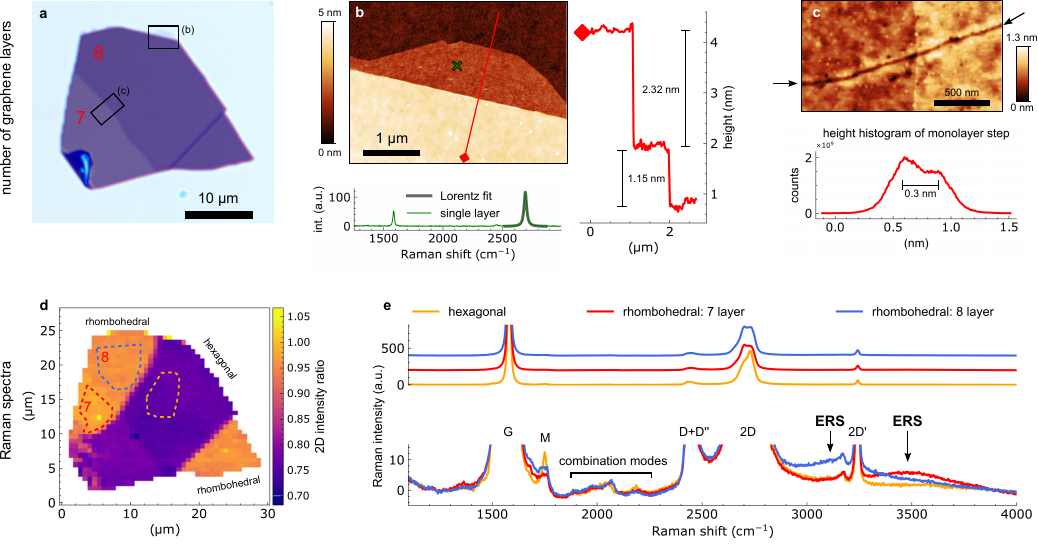}
	\caption{\textbf{Raman spectra of rhombohedral graphite.}
		\textbf{(a)} Optical microscopy image of a graphite flake.
		Red numbers indicate the number of graphene layers.
		Black rectangles show the positions of the AFM images in (b) and (c).
		\textbf{(b)} AFM (tapping mode) topography image of the flake in (a). The flake has a single graphene layer protrusion, which was used as the reference for AFM height measurements.
		The single layer nature of the flake is shown by the green Raman spectrum in the lower inset, measured at the position shown by the green "x".
		Right inset: height section of the flake along the red line.
		The flake is $2.32$ nm thick, relative to the bottom graphene, meaning 8 graphene layers in total.
		\textbf{(c)} AFM (tapping mode) topography image of the single layer step in the middle of the flake.
		The hexagonal - rhombohedral domain wall is marked by the black arrows.
		Lower inset shows the height histogram of the image.
		\textbf{(d)} Integrated intensity ratio of the 2D peak.
		\textbf{(e)} Raman spectra averaged in the areas marked by correspondingly coloured dashed outlines in (d), each spectrum is an average of 50 to 80 spectra, with an individual integration time of 2 s.
		Top panel: spectra are offset for clarity.
		Bottom panel: same spectra as in the top panel, showing the background signal.
		Electronic Raman scattering (ERS) peaks marked by black arrows.
		Raman spectra are measured, using 488 nm excitation.
	}
	\label{fig:id}
	\end{center}
\end{figure}

\begin{figure}[!ht]
	\begin{center}
	\includegraphics[width = 1 \textwidth]{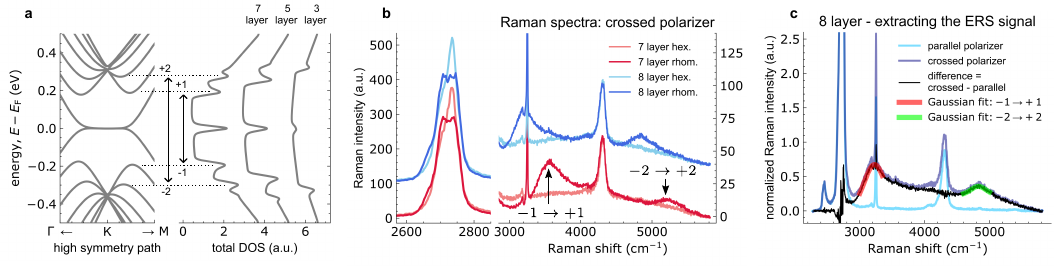}
	\caption{\textbf{Extracting the ERS signal.}
		\textbf{(a)} Left: \emph{ab initio} band structure around the K point of 7 layer RG.
		Energy is with respect to the Fermi level ($E_{\mathrm{F}}$).
		Right: density of states (DOS) at selected RG thicknesses.
		The transitions between the DOS peaks, which result in the ERS signal, are shown by arrows.
		\textbf{(b)} Raman spectra of the 7 and 8 layer regions in Fig.~\ref{fig:id}, measured using crossed polarization.
		Arrows mark the ERS signal, associated with the transitions between the DOS peaks.
		\textbf{(c)} Example of extracting the ERS signal.
		This is achieved by subtracting the spectrum measured with parallel polarization from the one acquired using the crossed polarizer configuration.
		Prior to subtraction, both spectra are normalized to the 4300 cm$^{-1}$ peak. 
		Gaussian fits applied to the resultant ERS signals are also displayed.
		Raman spectra are measured, using 488 nm excitation, integration time for each spectrum is 20 s.
	}
	\label{fig:ers_extracting}
	\end{center}
\end{figure}

Although the 2D peak shape converges with large $N$, its integrated intensity ratio should be used as a first tool to rapidly screen graphite flakes for RG containing flakes.
Furthermore, determining the precise number of graphene layers in the flake is a crucial step in identifying RG without stacking faults, therefore we describe this step in detail below.
The layer number can be easily determined by optical contrast~\cite{Blake2007-kg,Uslu2023-zs}, in the case of <10 layers.
For thicker samples atomic force microscopy (AFM) can be used, provided care is taken during the measurement, to counter any anomalous height signal stemming from the difference in properties of the graphite and substrate~\cite{Nemes-Incze2008-la}.
Fig.~\ref{fig:id}a-c illustrates the thickness measurement of a graphite flake exfoliated onto a Si / SiO$_2$ substrate, where a graphene layer slightly extends near the top in the optical microscopy image (Fig.~\ref{fig:id}a).
For accurate height measurement using dynamic atomic force microscopy (AFM) methods, it is essential to measure the flake's thickness relative to this bottom graphene layer, or in an area where the flake folds back onto itself~\cite{Nemes-Incze2008-la}.
Raman spectroscopy confirms that the protruding graphene is indeed a single layer.
The well-known height anomaly in dynamic (tapping mode) AFM images~\cite{Nemes-Incze2008-la} is evident in Fig.~\ref{fig:id}b.
Here, the graphene layer's height, relative to the SiO$_2$ substrate, registers at 1.15 nm instead of the expected 0.33 nm.
Conversely, the single-layer step in the middle of the flake is accurately measured, aligning with the van der Waals distance of graphite, as illustrated in Fig.~\ref{fig:id}c.
Therefore, when measured relative to the bottom graphene layer, the flake's thickness is determined to be 2.32 nm, corresponding to seven graphene layers.
Including the bottom layer, our flake comprises regions with eight and seven layers, respectively.
The layer number is also reproduced from optical contrast measurements of the layer thickness.

The AFM topography image in Fig.~\ref{fig:id}c reveals a domain wall within the flake.
A Raman spectroscopy map measured across this flake (Fig.~\ref{fig:id}d) shows that the domain wall separates areas with wide and narrow 2D peaks.
In both the 7 and 8 layer regions, the larger 2D peak intensity ratio suggest rhombohedral stacking.
Fig.~\ref{fig:id}e features averaged Raman spectra from the hexagonal, and the 7 and 8 layer rhombohedral regions.
The bottom panel focuses on the Raman spectra's background signal.
In addition to the well known low-intensity phonon modes (M, iTALO, iTOTA, LOLA),~\cite{Kawashima1995-bd,Tan2001-vp,Nguyen2014-nu} the background reveals a wide peak on both the 7 and 8 layer rhombohedral areas.
The halfwidths of these peaks is in the 500 cm$^{-1}$ regime, consistent with the temperature broadening at room temperature  ($3 k_{\mathrm{B}} T = 77.5$ meV $\approx 625\ \mathrm{cm}^{-1}$ for $T = 300$ K, where $k_{\mathrm{B}}$ is the Boltzmann constant).
Their absence in the hexagonal region implies that their origin is electronic Raman scattering, because the ERS in hexagonal graphite is mostly flat~\cite{Garcia-Ruiz2019-dp,McEllistrim2023-op} due to a lack of a bulk gap and other sharp DOS features.

The energy of the RG band edges varies with $N$, as illustrated by Fig.~\ref{fig:ers_extracting}a, which shows the density of states (DOS) for 3, 5, and 7-layer RG from \emph{ab initio} calculations detailed in the Methods.
In the DOS, we observe peaks labelled $+1$, $+2$ for unoccupied states and $-1$, $-2$ for occupied states.
Notably, the energy gap between the $+$ and $-$ peaks narrows as the number of rhombohedrally stacked graphene layers increases.
The transitions between peaks $-1 \rightarrow +1$ and $-2 \rightarrow +2$ dominate in the electronic Raman scattering (ERS)~\cite{Garcia-Ruiz2019-dp,McEllistrim2023-op}.
The marked layer-number dependency of these energy separations makes the ERS signal a unique fingerprint of perfectly stacked RG if the layer number is known from AFM or optical contrast measurements.
As is obvious from Fig.~\ref{fig:id}e, the signal is $\sim$1\% of the 2D peak intensity.
This makes it undetectable in most measurements, where the integration time is insufficient to resolve the smallest of peaks.
We can substantially improve the relative intensity of the ERS with respect to the phonon peaks, by using a polarizer in the path of the scattered beam.
Of course in this case a somewhat longer integration time is needed.

A polarization dependent measurement can also be used to verify that the broad peaks observed in our sample originate from ERS.
If the incoming excitation laser is linearly polarized, the optical selection rules of the ERS process result in a perpendicular linear polarization of the scattered photons~\cite{Kashuba2012-ae,Riccardi2016-ir,Garcia-Ruiz2019-dp}.
Fig.~\ref{fig:ers_extracting}b displays the 2D peak alongside the broad ERS response for the 7 and 8 layer regions of our flake.
Unlike the spectrum in Fig.~\ref{fig:id}e, a polarizer is now positioned in the path of the scattered light, oriented perpendicularly to the polarization of the incoming excitation laser.
This "crossed polarizer" setup amplifies the relative visibility of the ERS signal, which becomes $\sim$10\% of the 2D peak intensity, as shown in Fig.~\ref{fig:ers_extracting}b.
Consequently, we can confirm that the broad peaks stem from ERS by comparing measurements from the same location under both crossed and parallel polarizer configurations, as depicted in Fig.~\ref{fig:ers_extracting}c.
In the parallel configuration, the broad peaks are absent, whereas they are clearly present in the crossed polarizer setup.
This comparison can also be used to separate the ERS peaks from the phonon peaks~\cite{Kawashima1995-bd,Tan2001-vp}.
By performing the same measurement in the crossed and parallel configuration we can subtract the two curves, which leaves us with the ERS signal.
We attribute the peak exhibiting the lower Raman shift to the $-1 \rightarrow +1$ transition, and the peak at higher shift is identified as the $-2 \rightarrow +2$ process.
To extract the ERS signal, we first subtract a linear background and normalize to a phonon peak, in this example (Fig.~\ref{fig:ers_extracting}c) to the 2D + G peak at 4300 cm$^{-1}$.
After this we subtract the two spectra, which leaves us with the ERS signal.
We fit the resulting ERS peaks by Gaussians to determine their Raman shift.
The result of this procedure is shown in Fig.~\ref{fig:ers_extracting}c.
A detailed description of the fitting process can be found in section S6 of the Supplementary Information and the shared data~\cite{Nemes-Incze_Peter_Topology_in_Nanomaterials_Research_Group2024-kj}.

\begin{figure}[!ht]
	\begin{center}
	\includegraphics[width = 1 \textwidth]{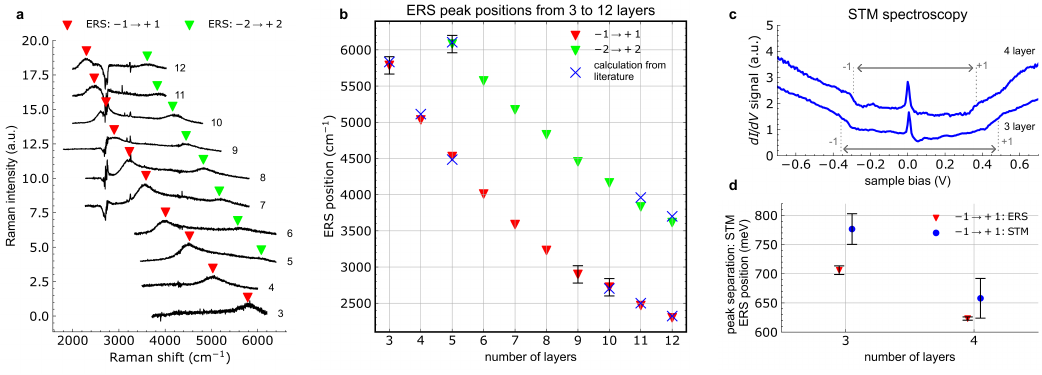}
	\caption{\textbf{Layer number dependence of ERS peaks in RG.}
		\textbf{(a)} Difference of crossed/parallel spectra for graphene layer numbers between 3 and 12.
		Positions of the ERS peaks are shown by red and green triangles.
		From 3 to 8 layers, the spectra are normalized to the 2D + G mode at 4300 cm$^{-1}$.
		For 9 layers, the 2D peak and for 10, 11 and 12 layers the 2D' (3247 cm$^{-1}$) peak was used for normalization.
		Spectra are offset and scaled along the y axis for better visibility.
		\textbf{(b)} ERS peak positions for the first and second transitions.
		Blue crosses show the calculated ERS peak positions from refs~\cite{Garcia-Ruiz2019-dp,McEllistrim2023-op}.
		Error bars that are not shown are smaller than the symbols (40 cm$^{-1}$).
		\textbf{(c)} Direct measurement of the DOS, for 3 (bottom) and 4 layer (top) RG by scanning tunnelling microscopy (STM).
		The energy gap associated with the $-1 \rightarrow +1$ transitions is shown by grey arrows.
		\textbf{(d)} Comparison of the direct DOS peak energy separation (from STM) and the ERS signal measured on 3 and 4 layers.
		Red data points denote the average from measurements across multiple flakes: four for trilayer and three for tetralayer.
		Error bars for Raman measurements represent weighted estimated errors, while STM error bars are based on the standard deviation of energy separation values across the sample.
		Raman spectra are measured, using 488 nm excitation.
		STM data was measured at a temperature of 9 K.
	}
	\label{fig:ers_in_rg}
	\end{center}
\end{figure}

We apply this procedure to samples with thicknesses varying from 3 to 12 graphene layers.
Each sample undergoes meticulous examination using atomic force microscopy (AFM) to ensure accurate layer count.
Additionally, we identify few layer samples through optical contrast measurements~\cite{Blake2007-kg,Uslu2023-zs}.
The ERS peaks of these samples can be seen in Fig.~\ref{fig:ers_in_rg}a.
We can observe a clear trend: with increasing layer number the $-1 \rightarrow +1$ and $-2 \rightarrow +2$ peaks continuously shift to lower energy (lower Raman shift).
By plotting these peak positions extracted by Gaussian fitting versus the layer number this trend is even more evident (see Fig.~\ref{fig:ers_in_rg}b).
In certain cases the ERS peak positions have a larger estimated error ($\pm$120 cm$^{-1}$) due to lower signal to noise or due to the overlap of the $-1 \rightarrow +1$ ERS peak with the 2D mode in the case of 9 and 10 layers.
These measurements align within our error margins with the calculated positions of ERS peaks from references:~\cite{Garcia-Ruiz2019-dp,McEllistrim2023-op} (see Fig.~\ref{fig:ers_in_rg}b).
The measured ERS peak positions are also supplied as a table in the Methods (see Table~\ref{table:ers}).

The ERS measurements appear to directly reveal the energy of the electron-hole excitations related to the DOS peak separations ($-1 \rightarrow +1$ and $-2 \rightarrow +2$).
Nonetheless, the measured ERS energies can differ from energies between the DOS peaks due to many–body excitonic~\cite{Quintela2022-yy} and polarization effects~\cite{Tepliakov2021-na} originated from the external electric field of the laser, among others.
To assess these factors, we directly measure the surface DOS of trilayer and tetralayer sample surfaces using scanning tunnelling microscopy (STM) at a temperature of 9 K.
Comparisons between STM and ERS measurements are valid as the $\pm 1$ and $\pm 2$ peaks occupy identical positions in both the total (probed by ERS) and surface (probed by STM) DOS.
In STM spectroscopy, the DOS peaks depicted in Fig.~\ref{fig:ers_extracting}a manifest as distinct shoulders in the tunnelling conductance, as shown in Fig.~\ref{fig:ers_in_rg}c.
We have quantified the separation between the $-1$ and $+1$ shoulders and tracked its variation across the sample surface (see section S7 of the Supplementary Information).
The standard deviation of this variation is represented as the error bar in Fig.~\ref{fig:ers_in_rg}d.
For trilayer samples, a notable discrepancy beyond the error margins is observed in both Raman and STM measurements, suggesting that ERS peak energies are 70$\pm$27 meV lower than the DOS peak spacings.
Consequently, the collective influence of excitonic and other effects on the ERS process is estimated to be in the tens of meV range for trilayer, and lower for tetralayer.
We note that this value is much smaller than the difference between the measured 3 and 4 layers ERS peak position values, thus do not affect the identification of the few layer RG samples.

\begin{figure}[h]
	\begin{center}
	\includegraphics[width = 1 \textwidth]{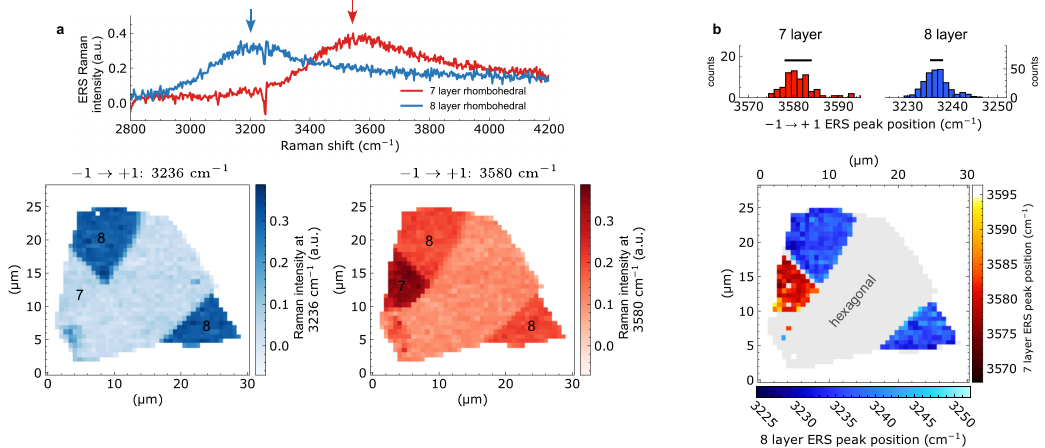}
	\caption{\textbf{Mapping the ERS across a flake.}
		\textbf{(a)} Intensity of the ERS signal across the 7 and 8 layer flake.
		Top: selected spectra, where the phonon peaks are removed as shown in Fig.~\ref{fig:ers_extracting}c.
		Coloured arrows show the Raman shift, for which the ERS intensity is plotted in the bottom panel.
		Bottom: maps of the ERS intensity for the $-1 \rightarrow +1$ transition for the 8 (blue) and 7 layer (red) ERS peak.
		Black numbers show the number of graphene layers in the region.
		\textbf{(b)} ERS peak position in cm$^{-1}$ for the 7 and 8 layer areas, as determined by Gaussian fitting (see Fig.~\ref{fig:ers_extracting}c).
		The whole flake is marked as the grey area.
		Top: histograms of the peak positions, black bars show the size of the standard deviation (7 layer: 6 cm$^{-1}$, 8 layer: 2.9 cm$^{-1}$).
		Raman spectra are measured, using 488 nm excitation.
	}
	\label{fig:ers_map}
	\end{center}
\end{figure}

Next we check the homogeneity of the ERS peak position over the flake.
Taking the difference in the crossed and parallel polarization measurements across the crystal, we plot the Raman intensity of the $-1 \rightarrow +1$ transition for the 8 and 7 layer areas.
It becomes clear that the largest intensity of the ERS peak is located exactly on the rhombohedral 7 and 8 layer regions  respectively (Fig.~\ref{fig:ers_map}a).
To check the homogeneity of the rhombohedral regions, in Fig.~\ref{fig:ers_map}b we plot the position of the ERS peak.
The standard deviation within both regions is under 5 cm$^{-1}$.
If a stacking fault were present in these regions, a change in peak energy two orders of magnitude greater would be expected.
This is because the shift in ERS peak position, due to a stacking fault, is of the order of the change in energies with adding another layer to perfectly stacked RG~\cite{Garcia-Ruiz2019-dp}.
Thus, the change in peak position would be least 300 cm$^{-1}$ in the presence of a stacking fault.

\begin{figure}[h]
	\begin{center}
	\includegraphics[width = 1 \textwidth]{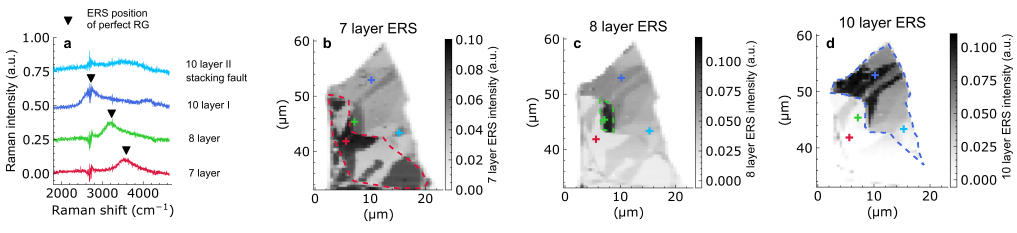}
	\caption{\textbf{Identifying perfect rhombohedral domains and stacking faults.}
		\textbf{(a)} ERS signal extracted from the spectra shown in Fig.~\ref{fig:problem}a.
		Spectra are offset for clarity.
		The top spectrum shows a shallow ERS peak around the position of the 7 layer, implying that the area contains a stacking fault.
		\textbf{(b)} Map of the average ERS intensity in the 3350 cm$^{-1}$ to 3650 cm$^{-1}$ range.
		\textbf{(c)} Map of the average ERS intensity in the 3000 cm$^{-1}$ to 3300 cm$^{-1}$ range.
		\textbf{(d)} Map of the average ERS signal intensity in the 2400 cm$^{-1}$ to 2900 cm$^{-1}$ range.
		The outline of the regions with 7, 8 and 10 graphene layers are marked by the correspondingly coloured dashed lines.
		The measurement positions of the spectra in (a) are shown by correspondingly coloured "+" marks.
		These maps were extracted from the same data as shown in Fig.~\ref{fig:problem}, the spectra in (a) are from the same position as in Fig.~\ref{fig:problem}.
		Raman spectra are measured, using 532 nm excitation.
	}
	\label{fig:stacking_fault}
	\end{center}
\end{figure}

Stacking faults in the sample can be identified in one of two ways.
Since some stacking faults don't have strong peaks in the DOS~\cite{McEllistrim2023-op}, they can be identified, by the lack of an ERS peak in a region with "RG-like" 2D peak shape.
An example of this is the spectrum shown in Fig.~\ref{fig:stacking_fault}a, marked "10 layer II".
The 2D peak of this spectrum is displayed in Fig.~\ref{fig:problem}a.
Alternately, if the Raman shift of the ERS peak does not match the expected value for the given layer number, we have a clear signature of a stacking fault.
To show the necessity of the ERS measurement, beyond the determination of the 2D peak shape, compare the data in Fig.~\ref{fig:problem} and Fig.~\ref{fig:stacking_fault}.
Both figures stem from the same Raman map.
Both of the two (light and dark) blue spectra are measured on the 10 layer region of the flake, having very similar 2D peak shapes (see Fig.~\ref{fig:problem}a).
However, of we plot the ERS peaks of the two spectra it becomes clear that the spectrum marked "10 layer I", has defect free rhombohedral stacking, while the other one ("10 layer II") shows only a weak ERS signal, above the expected Raman shift (see Fig.~\ref{fig:stacking_fault}a).
Plotting the Raman intensity of the ERS signal at the ERS peaks for perfect stacking, we can map the areas in the flake which have no stacking faults (see Fig.~\ref{fig:stacking_fault}b, c, d).
In the map of the 10 layer ERS (Fig.~\ref{fig:stacking_fault}d), we can clearly discern a stacking fault running across the perfect rhombohedral stacking, which is barely visible in the 2D peak shape map (Fig.~\ref{fig:problem}b).
While identifying the exact structure of the stacking faults is possible in principle, it requires specific calculations of the band structure and ERS signal for each stacking sequence, the number of which increases exponentially with $N$~\cite{Atri2023-te}.

\section*{Conclusions}


The introduction of ERS as a fast and accessible optical characterization method to identify rhombohedral graphite without stacking faults, breaks down a major hurdle in exploring the properties of RG.
Our results emphasize the significance of measuring the ERS spectrum in thicker RG samples, in addition to their phonon peaks, to uncover or rule out hidden stacking faults.
These faults could lead to substantial discrepancies in results obtained from seemingly similar samples.
This is helped by the fact that Raman spectroscopy is already a widely used characterization tool in studying van der Waals materials.
ERS characterization is expected to  enable a consistent comparison of results form different samples and research groups, which is essential for the development of the field.
Moreover, we expect that ERS can be extended to the identification of various stacking faults in RG, some of which are predicted to harbour unique properties~\cite{Garcia-Ruiz2023-tu}, including ferroelectricity~\cite{Garcia-Ruiz2023-so}.

\subsection*{Methods}

\subsubsection*{Exfoliation and AFM characterization of samples}

We exfoliate samples using "blue tape" (Ultron systems P/N: 1008R-8.0), but other tapes work equally well.
As substrate we used Si wafers (90 nm SiO$_2$).
Graphite samples were purchased from NGS Trading \& Consulting GmbH (www.graphit.de).
Based on hundreds of flakes investigated by Raman measurements, roughly 40\%\ of them have some rhombohedral domains.
We performed AFM measurements, using an NX10 microscope from Park Systems in tapping mode (non-contact mode).

\subsubsection*{Raman measurements}

For Raman measurements we use a Witec 300rsa+ confocal Raman system, using 488, 532, and 633 nm laser excitations.
All data shown in the main text are measured using a diffraction grating with 600 lines $\cdot$ mm$^{-1}$.
Raman measurements are first analysed using the Witec data processing software supplied with the confocal Raman system.
We performed final data processing and generated the figures, using the open-source Python tool: Ramantools~\cite{Nemes-Incze2023-ea}.

\begin{table}[h]
	\centering
	\begin{tabular}{c|c|c}
		 Layer Number & $-1 \rightarrow +1$ ERS peak position (cm$^{-1}$) & $-2 \rightarrow +2$ ERS peak position (cm$^{-1}$) \\
		 \hline
		 \hline
		            3 & 5785 $\pm$ 120 & \\
		            \hline
		            4 &  5032 $\pm$ 40 & \\
		            \hline
		            5 &  4526 $\pm$ 40 & 6080 $\pm$ 120 \\
		            \hline
		            6 &  4008 $\pm$ 40 & 5571 $\pm$ 40 \\
		            \hline
		            7 &  3586 $\pm$ 40 & 5172 $\pm$ 40 \\
		            \hline
		            8 &  3228 $\pm$ 40 & 4827 $\pm$ 40 \\
		            \hline
		            9 & 2898 $\pm$ 120 & 4453 $\pm$ 40 \\
		            \hline
		           10 & 2721 $\pm$ 120 & 4163 $\pm$ 40 \\
		           \hline
		           11 &  2467 $\pm$ 40 & 3830 $\pm$ 40 \\
		           \hline
		           12 &  2299 $\pm$ 40 & 3613 $\pm$ 40 \\
		\hline
	\end{tabular}
	\caption{Table of measured ERS peak positions, as shown in Fig.~\ref{fig:ers_in_rg}.}
	\label{table:ers}
\end{table}

The largest contribution to the error bars in the ERS peak position are determined by the fitting window chosen for the Gaussian fit.
The experimental variability of the ERS peak positions is much lower than this, see Fig.~\ref{fig:ers_map}b.

\subsubsection*{STM measurements}

STM measurements are carried out at a temperature of 9 K, using an instrument from RHK (PanScan Freedom), with a base pressure of 5$\times$10$^{-11}$ Torr.
Tunnelling conductance measurements were performed, using a Lock-in amplifier, at a frequency of 1372 Hz and a bias modulation amplitude of 5.5 mV.
Samples investigated by STM are exfoliated flakes, supported on a Si/SiO$_2$ substrate and contacted using In spikes~\cite{Girit2007-vi}.
Data analysis of the STM measurements is carried out using the open-source Python tool: RHKPY~\cite{Nemes-Incze2023-fd}.

A potential systematic error of $\sim$10 meV could exist in determining the $-1 \rightarrow +1$ gap in STM.
The band edges have a step and a peak in the calculated DOS, the peaks themselves have the largest contribution to the ERS signal~\cite{McEllistrim2023-op}.
However, in the case of the STM measurement, we don't observe any peaks only a step in the tunnelling conductance.
In evaluating the $-1 \rightarrow +1$ gap in our STM measurements, we fit the top of this step, as shown by the dashed vertical lines in Fig.~\ref{fig:ers_in_rg}c (see section S7 Supplementary Information for details).
Due to this effect, there might still be an additional error of $\sim$10 meV in the gap size, as measured by STM.
To make better measurements of the surface DOS, lower temperature STM measurements are needed to be able to resolve the peak itself at the DOS step.

\subsubsection*{Calculation details}

The optimized geometry and ground state Hamiltonian and overlap matrix elements of each structure were self consistently obtained by the SIESTA implementation of density functional theory (DFT)~\cite{artacho2008siesta,soler2002siesta,garcia2020siesta}.
SIESTA employs norm-conserving pseudopotentials to account for the core electrons and linear combination of atomic orbitals to construct the valence states.
For all cases the considered samples were separated with a minimum of 1.35 nm thick vacuum in the perpendicular direction.
The generalized gradient approximation of the exchange and the correlation functional was used with Perdew-Burke-Ernzerhof parametrization~\cite{perdew1996generalized} with a double-$\zeta$ polarized basis set.
The geometry optimizations were performed until the forces were smaller than 0.1 eV nm$^{-1}$ .
The geometry of the considered structures were optimized for every configuration, initiated from the experimental in-plane lattice constant $a$ = 0.246 nm and out-of-plane lattice constant $c$ = 0.670 nm of hexagonal graphite.
During the geometry relaxation the real-space grid was defined with an equivalent energy cutoff of 400 Ry and the Brillouin zone integration was sampled by a 120$\times$120$\times$1 Monkhorst-Pack $k$-grid \cite{monkhorst1976special}.

\subsection*{Acknowledgements}
	
	Funding from the National Research, Development, and Innovation Office (NKFIH) in Hungary, through the Grants K-146156, PD-146479, K-134258, K-132869, FK-142985, \'{E}lvonal KKP 138144, 2022-1.2.5-T\'{E}T-IPARI-KR and TKP2021-NKTA-05 are acknowledged.
	PNI and AP acknowledge support from the J\'{a}nos Bolyai Research Scholarship of the Hungarian Academy of Sciences.
	KK was supported by the \'{U}NKP-23-3-II-BME-161 New National Excellence Program of the Ministry for Culture and Innovation from the source of the National Research, Development and Innovation Fund.
	TZ and GM acknowledge financial support from Slovak Academy of Sciences project IMPULZ IM-2021-42, FLAG ERA JTC 2021 2DSOTECH, and Slovak Research and Development Agency provided under Contract No. APVV-SK-CZ-RD-21-0114.
	The authors thank Markus Morgenstern and Aitor Garc\'{i}a-Ruiz for fruitful discussions.

\subsection*{Author contributions}

	KM was responsible for preparing the samples.
	The Raman spectra and maps were measured by KM and AP.
	KK conducted the STM measurements.
	Data analysis was performed by AP, KM, LT and PNI.
	ZT, VP, and MG carried out \emph{ab initio} calculations.
	The project was conceived and coordinated by PNI, with support from AP.
	PNI wrote the manuscript with contributions from all authors.

\subsection*{Data availability} 

	The data that support the findings of this study are openly available at Zenodo at DOI: 10.5281/zenodo.10931524, reference number~\cite{Nemes-Incze_Peter_Topology_in_Nanomaterials_Research_Group2024-kj}.

\subsection*{Competing interests}

	The authors declare no competing interests.


\newpage


\end{document}